\documentclass[11pt, a4paper]{article}
\usepackage[margin=2cm]{geometry}
\usepackage{times,graphicx,amsmath,amssymb,pdflscape,natbib,authblk,lineno,setspace,rotating}
\usepackage[lofdepth,lotdepth]{subfig}
\usepackage{hyperref}

\newcommand{\Div}{{\boldsymbol{\nabla}}\cdot}
\newcommand{\Grad}{{\boldsymbol{\nabla}}}

\newcommand{\diff}[2]{\frac{\textrm{d}{#1}}{\textrm{d}{#2}}}
\newcommand{\pdiff}[2]{\frac{\partial{#1}}{\partial{#2}}}

\newcommand{\strr}{\dot{\varepsilon}}
\newcommand{\Strr}{\dot{\boldsymbol{\varepsilon}}}
\newcommand{\xvec}{\boldsymbol{x}}
\newcommand{\gvec}{\boldsymbol{g}}
\newcommand{\xhat}{\hat{\boldsymbol{x}}}
\newcommand{\zhat}{\hat{\boldsymbol{z}}}
\newcommand{\vs}{\boldsymbol{v}_s}
\newcommand{\vl}{\boldsymbol{v}_\ell}
\newcommand{\dflux}{\boldsymbol{q}}
\newcommand{\vbar}{\overline{\boldsymbol{v}}}

\newcommand{\omp}{(1-\phi)}
\newcommand{\mobility}{K_\phi}
\newcommand{\permexp}{r}

\title{Magmatic focusing to mid-ocean ridges: the role of grain size
  variability and non-Newtonian viscosity}
\author{Andrew J.~Turner$^0$, Richard F.~Katz$^1$, Mark Behn$^2$ \&
  Tobias Keller$^3$}
\affil{$^0$ AWE, Aldermaston, Berkshire, UK; 
  $^1$ Department of Earth Sciences, University of Oxford, Oxford,
  UK; $^2$ Department of Geology and Geophysics, Woods Hole 
  Oceanographic Institution, Woods Hole, Massachusetts, USA; 
  $^3$ Department of Geophysics, Stanford University, Stanford, USA.}
\affil{email: richard.katz@earth.ox.ac.uk}

\begin{document}
\maketitle
\doublespace

\begin{abstract}
  Melting beneath mid-ocean ridges occurs over a region that is much
  broader than the zone of magmatic emplacement to form the oceanic
  crust. Magma is focused into this zone by lateral transport. This
  focusing has typically been explained by dynamic pressure gradients
  associated with corner flow, or by a sub-lithospheric channel
  sloping upward toward the ridge axis.  Here we discuss a novel
  mechanism for magmatic focusing: lateral transport driven by
  gradients in compaction pressure within the asthenosphere.  These
  gradients arise from the co-variation of melting rate and compaction
  viscosity. The compaction viscosity, in previous models, was given
  as a function of melt fraction and temperature.  In contrast, we
  show that the viscosity variations relevant to melt focusing arise
  from grain-size variability and non-Newtonian creep.  The
  asthenospheric distribution of melt fraction predicted by our models
  provides an improved explanation of the electrical resistivity
  structure beneath one location on the East Pacific Rise.  More
  generally, although grain size and non-Newtonian viscosity are
  properties of the solid phase, we find that in the context of
  mid-ocean ridges, their effect on melt transport is more profound
  than their effect on the mantle corner-flow.
\end{abstract}

\section{Introduction}

Melting beneath mid-ocean ridges occurs over a broad region, of order
100~km width in the plate-spreading direction.  By contrast, magmatic
emplacement to form the oceanic crust is limited to a zone that is
$\sim$5~km in width, centered on the ridge axis.  Magmatic focusing
refers to the process or processes by which magma is extracted from
the broad source region and transported to the narrow emplacement
zone.  The mechanism, lateral extent, and efficiency of magmatic
focusing have been a subject of considerable interest in theoretical 
models of mid-ocean ridge magmatism.

There are two dominant hypotheses for magmatic focusing in that
literature. The first argues that the dynamic pressure gradient
arising from plate-driven, asthenospheric corner-flow sucks melt
laterally toward the ridge axis \citep{Spiegelman87,
  Phippsmorgan87}. The second argues that crystallisation at the base
of the thermal lithosphere creates an impermeable layer sloping upward
toward the ridge axis; magma is then channelised along this barrier
toward the ridge axis \citep{Sparks91,Spiegelman93b,Hebert10}. This
latter mechanism has been used to model melt focusing in 3-D along
specific ridge systems \citep[e.g.,][]{Magde97,Montesi11} and oceanic
transform faults \citep{Gregg09,Bai15}. \cite{Keller17} describe
two-phase flow models indicating that both mechanisms contribute to
focusing.  However, it is unclear whether either corner-flow suction
or the sublithospheric-permeability-barrier mechanism are supported by
observations. To focus melt over distances of $\sim$60~km, the former
requires asthenospheric viscosities ($\sim$10$^{21}$~Pa-sec) that are
unrealistically high, based on laboratory estimates
\citep{Hirth03}. The latter predicts a high-porosity layer (melt
volume fractions of several percent) beneath the lithosphere that
should be detectable by seismic or magnetotelluric methods, but to
date has not been detected in the vicinity of a mid-ocean ridge.
Indeed, a recent magnetotelluric inversion of electrical resistivity
beneath the northern East Pacific Rise suggests a style of magmatic
flow than has not been predicted by any models of melt focusing at
mid-ocean ridges \citep{Key13}.

\cite{Key13} image a triangular region of diminished resistivity
between about 10 and 100~km depth beneath the ridge axis. The sides of
this region are relatively sharp and dip at about 45$^\circ$ away from
the ridge axis. Assuming that this resistivity structure is broadly
representative of the volume fraction of silicate melts, this
observation is inconsistent with the
sublithospheric-permeability-barrier mechanism for melt focusing:
there is no evidence for an off-axis, low-resistivity layer at depths
immediately beneath the thermal lithosphere.  Rather, there is a
high-resistivity region that extends off-axis from $\sim$5~km depth to
the sides of the low-resistivity triangle.  Previous focusing
hypotheses \citep{Spiegelman87, Sparks91} and more recent
computational models \citep{Ghods00, Katz08, Keller16} predict
non-zero or even upward-increasing melt fraction in these regions,
which should be associated with modest to low resistivity
\citep{Miller15}.

These observations motivate a re-examination of the mechanisms by
which melt is focused beneath mid-ocean ridges. Permeability, which is
a key control on melt segregation, is highly sensitive to grain size
\citep{Miller14, VBW86}. As such, several studies have analysed the
effects of grain-size variability on magmatic focusing in subduction
zones \citep{cagnioncle07, Wada11, Wada15}. For mid-ocean ridges,
\cite{Turner15} coupled single-phase, passive mantle flow with
composite, non-Newtonian viscosity and grain size evolution.  Steady
state solutions predicted an upward decrease in grain size within an
elongated zone that slopes toward the ridge axis. Although the model
by \cite{Turner15} did not include an explicit prediction of magmatic
flow, the layer of small grain size was interpreted as a barrier to
vertical magmatic segregation (albeit a leaky one). Intriugingly, this
layer is coincident with the margin of the low-resistivity region in
the \cite{Key13} MT observations. This led \cite{Turner15} to
hypothesise that grain size plays an important role in focusing magma
toward the axis at mid-ocean ridges.

In the present manuscript, we develop two-phase flow models of
magma/mantle interaction in a variable grain-size mantle to
investigate the focusing hypothesis of \cite{Turner15}.  Our models
compute the mean grain-size field as in that work, but further couple
it with conservation of mass, momentum, and energy for a liquid phase
(the magma) and a solid phase (the mantle) \citep{McKenzie84,
  Keller16}.  We obtain a pattern of magmatic segregation that
exhibits strong focusing toward the ridge axis. However, the focusing
mechanism is more complex than envisioned by \cite{Turner15} and
cannot be explained by the spatial structure of permeability that
arises from the grain-size field.  In particular, focusing is linked
to variations in melting rate and asthenospheric viscosity; it is
linked to grain size, but through its influence on viscosity. The
resulting focusing mechanism is previously unrecognised and predicts
melt distributions beneath the ridge axis that are broadly consistent
with the resistivity structure of \cite{Key13}.  Below we present our
numerical two-phase results after a brief exposition of theory and
methods used in this study.  These results are followed by a
discussion of the causative physical mechanisms for melt focusing.

\section{Methods}

The overall method used in this work is computational modelling of
geodynamic processes by solution of partial differential equations. In
this section we describe the physical processes that are incorporated
in the governing equations and the assumptions that are used to
simplify those equations.  We provide a brief overview of the
numerical methods used to discretise and solve the system.

\subsection{Mechanics and thermochemistry}

The two-phase model comprises statements of conservation of mass and
momentum for a liquid ($\ell$) and a solid ($s$) phase. The volume
fraction of liquid within a representative volume element is
$\phi(\xvec,t)$. The density of each phase is considered constant and
the two densities are considered equal (to $\rho$) except in terms
representing the body force, where the solid-minus-liquid difference
is denoted $\Delta\rho$ (the extended Boussinesq
approximation). Following \cite{McKenzie84},
\begin{subequations}
  \label{eq:governing_mechanics}
  \begin{align}
    \label{eq:continuity}
    \Div\vbar &= 0,\\
    \label{eq:solid_mass}
    \pdiff{\phi}{t} - \Div\omp\vs &= \Gamma/\rho,\\
    \label{eq:stokes}
    \Grad P - \Div 2\eta\Strr - \Grad(\zeta_\phi-2\eta/3)\Div\vs
              &= -\phi\Delta\rho\gvec, \\
    \label{eq:darcy}
    \Grad P + \mobility^{-1}\phi(\vl-\vs)  &= -\Delta\rho\gvec.
  \end{align}
\end{subequations}
The first equation, in which $\vbar = \phi\vl+\omp\vs$, is a
continuity equation that expresses conservation of mass for the bulk,
two-phase system. The second equation represents conservation of mass
for the solid phase; $\Gamma$ is the melting rate in
mass/volume/time. The third equation represents Stokesian momentum
conservation for the aggregate;
$\Grad P \equiv \Grad P_\ell - \rho\gvec$ is a dynamic pressure
gradient;
$\Strr \equiv \tfrac{1}{2}\left[\Grad\vs + (\Grad\vs)^T\right]$ is the
strain-rate tensor; $\eta$ and $\zeta_\phi$ are the shear and
compaction viscosity, respectively; $\gvec$ is the accelleration of
gravity. The fourth equation represents Darcian momentum conservation
in the liquid phase; $\mobility\equiv k_\phi/\mu$ is the ratio of
permeability to magma viscosity.

The general characteristics and behaviour of this system of equations
is discussed in the literature.  Of particular interest here is the
Darcian segregation flux, $\dflux \equiv \phi(\vl-\vs)$, which is
driven by dynamic pressure gradients and magma buoyancy associated
with the interphase density difference, according to
equation~\eqref{eq:darcy}.

This system of equations is closed by equations for the melting rate
$\Gamma$, the liquid mobility $\mobility$ and the aggregate shear
$\eta$ and compaction $\zeta_\phi$ viscosities. The melting rate is
computed by coupling the mechanical system
\eqref{eq:governing_mechanics} with a thermochemical system that
represents conservation of energy and species mass, and with a kinetic
formulation for melting reactions called R\_DMC \citep[see][for
details]{Keller16}. We use a two-pseudocomponent system, where the
pseudocomponents are the product and residual of MORB-type mantle
melting \citep{Ribe85, Shorttle14}.

\subsection{Constitutive laws}

The permeability of the solid aggregate is parameterised on empirical
and theoretical grounds as $k_\phi = a^2\phi^\permexp/b$, where $a$ is the
mean grain size; $\permexp$ and $b$ are empirical constants, taken as
2 and 500 here \citep[][and refs.~therein]{Miller14}. This is valid at
porosities below the disaggregation threshold of about $0.3$. 

The viscosity of the aggregate is treated by \cite{Turner15} as a
harmonic mean of flow laws for diffusion creep, dislocation creep, and
dislocation-accommodated grain-boundary sliding (GBS). All of these
creep processes are thermally activated, but only dislocation creep
and GBS are non-Newtonian, and only diffusion creep and GBS are
grain-size sensitive.  To simplify the calculation of viscosity for
present purposes while retaining the generality to incorporate all of
these dependencies, we adopt a single shear-viscosity law,
\begin{equation}
  \label{eq:shear_viscosity}
  \eta =
  \eta_0\left<a\right>^{m/n}\left<\strr_{II}\right>^{(1-n)/n}
  \exp\left[\frac{E}{RT_0}\left(\frac{1}{\left<T\right>}-1\right)\right],
\end{equation}
where $\strr_{II}$ is the second invariant of the strain-rate tensor,
$m$ and $n$ are constants, $E$ is activation energy, $R$ is the
universal gas constant, and $T$ is temperature. Quantities in angular
brackets are normalised by a reference value, e.g.,
$\left<a\right> = a/a_0$. Variations in shear viscosity with melt fraction
are neglected because for small $\phi$, they are smaller than other
variations. See table~\ref{tab:symbols} for parameter and reference
values.

\begin{table}[ht]
  \centering
  \begin{tabular}{llll}
    Symbol & Description & Value & Units \\
    \hline
    $a$		& Grain size	& 0.003 or calculated	& m		\\
    $a_0$	& Reference grain size	&  $0.01$	& m		\\
    $\dot{\varepsilon}$		& Second invariant of strain rate	& calculated	& s$^{-1}$	\\
    $\dot{\varepsilon}_{II,0}$	& Reference second invariant of strain rate	& $2.5 \times 10^{-14}$	& s$^{-1}$\\
    $E$		& Activation energy  & $4.4 \times 10^{5}$	& J mol$^{-1}$		\\
    $T$		& Temperature	 & calculated	& K	\\
    $T_0$	& Reference temperature	& $1623$ & K	\\
    $\eta_0$ 	& Reference shear viscosity	& $5 \times 10^{18}$	& Pa s	\\
    $R_\zeta$	& Bulk viscosity constant	& $15$	& - 	\\
    $r$		& Permeability melt fraction exponent	& 2 & -	 \\
    $b$		& Permeability constant	& 500	& -		\\
    $p$		& Grain growth exponent		& 5	& -      \\
    $m$		& Grain rheology exponent	& See text  & -	\\
    $n$		& Stress exponent		& See text  & -	\\
    $K_g$ 	& Grain growth constant		& $6.67 \times 10^{-11}$	& m$^{5}$ s$^{-1}$	\\
    $E_g$	& Grain growth activation energy	& $3.35 \times 10^{5}$	& J mol$^{-1}$	\\
    $\psi$	& Surface energy density of grain boundaries	& $0.0625$	& J m$^{-2}$	\\
    $\rho$	& Density (Boussinesq)			& $3200$	& kg m$^{-3}$	\\
    $\Delta\rho$	& Density difference (Boussinesq) & $500$	& kg m$^{-3}$	\\
    $\phi$	& Melt fraction			& calculated	& -	\\
    $\Gamma$	& Melting rate			& calculated	& Kg m$^{-3}$ s$^{-1}$		\\
    $\mu$	& Liquid viscosity		& 10		& Pa s			\\
    $\vs$	& Solid velocity	& calculated	& m s$^{-1}$		\\
    $\vl$	& Fluid velocity	& calculated	& m s$^{-1}$	\\
    $\dflux$	& Separation flux	& calculated	& m s$^{-1}$	\\
    \hline
  \end{tabular}
  \caption{Meanings, values, and units for symbols used in this paper.}
  \label{tab:symbols}
\end{table}

The compaction viscosity is simply related to the shear viscosity
according to
\begin{equation}
  \label{eq:compaction_viscosity}
  \zeta_\phi = R_\zeta \frac{\eta}{\phi},
\end{equation}
where $R_\zeta$ is a dimensionless constant with a value between one
and twenty \citep{Takei09b, Simpson10a, Simpson10b}.

\subsection{Grain-size dynamics}

The permeability and the shear viscosity have an explicit dependence
on grain size, which is fundamental to the focusing hypothesis of
\cite{Turner15}. Hence we follow \cite{Turner15} in incorporating a
model of variations in the mean grain size based on the Wattmeter
formulation of \cite{Austin07,Austin09}. We give a brief review of the 
model and refer the reader to \cite{Turner15} and \cite{Behn09} for 
details.

The mean grain size varies according to
\begin{equation}
  \label{eq:grain-size}
  \pdiff{a}{t} + \vs\cdot\Grad a =
  \frac{K_g}{p}a^{1-p}\exp\left[-E_g/(RT)\right] - 
  \psi a^2\eta\Strr:\Strr.
\end{equation}
This equation states that grain-size variation along mantle flow lines
is due to independent processes of growth by material diffusion
between grains and reduction by recrystallisation. $K_g$ and $E_g$ are
the grain-growth prefactor and activation energy for grain growth,
respectively. $p$ is an exponent that, for an unpinned, single-phase
polycrystalline aggregate is in the range 2--3 \citep{BurTur52,
  Hill65, Atkin88}. Experiments suggest, however, that in the presence
of a minor, pinning phase, the grain-growth exponent should be taken
as 4--6 or even greater \citep{Hiraga10, Tasaka13, Thielmann15}. We
adopt the value of $p=5$ here.

The second term on the right-hand side of
equation~\eqref{eq:grain-size} represents grain-size reduction by
conversion of viscous work into formation of new grain
boundaries. $\psi$ is a coefficient representing the energetic cost of
grain boundaries. In models with composite viscosity, this term is
multiplied by a fraction representing the portion of deformational
work that is achieved by dislocation-based creep mechanisms.  for the
simplified, non-Newtonian flow law used here that fraction is unity,
but we assume that 25\% of the dissipated power goes into heating. 

\subsection{Boundary conditions and numerical methods}

The model is formulated as equations~\eqref{eq:governing_mechanics}
and \eqref{eq:grain-size} with constitutive
laws~\eqref{eq:shear_viscosity} and \eqref{eq:compaction_viscosity},
and with the thermo-chemical model of \cite{Keller16}. The body force
in the aggregate force balance equation~\eqref{eq:stokes} is of order
$\phi$ and hence we neglect it for simplicity. The domain is
two-dimensional, rectangular, and aligned vertically and with the
spreading direction. One lateral edge is directly beneath the
mid-ocean ridge axis and is assumed to be a line of bilateral
symmetry. The top boundary represents the surface of the oceanic crust
and has an imposed velocity of
$\vs(z=0) = (4\text{ cm/yr})\xhat\tanh(2x/x_r)$, where $x_r$ is the
width of distributed extension by normal faulting at the ridge
axis. The bottom (inflow) and side (outflow) boundaries have zero
normal stress gradient. The temperature on the top boundary is set to
zero Celcius, while the mantle potential temperature (on the bottom
boundary) is set to 1350$^\circ$C. Grain size has an imposed value of
3~mm on the bottom boundary. The fertile component is assumed to
comprise 20\% of the unmelted mantle source. Magma is allowed to leave
the domain at the ridge axis by enforcing no melting or freezing
within a region of width $x_r$ beneath the ridge axis. Over this
width, the pressure on the top boundary is set to enforce zero normal
gradient in the Darcy segregation flux.

The governing equations are discretised by finite volume and finite
difference methods on a staggered grid \citep{Katz07, Keller16} of
model dimensions 400~km wide by 250~km deep with grid resolution of
0.75~km. Time-stepping is semi-implicit and time-step size is computed
adaptively according to the magnitude of the maximum liquid velocity
over the domain. The system of discrete equations is solved at each
timestep with an outer iteration in which the mechanics, the
thermo-chemistry, and the grain size solutions are each updated
separately.  The mechanical solution is obtained using the sparse
direct solver MUMPS \citep{MUMPS:1}, while the other equations are
solved using block-Jacobi preconditioned GMRES. The data structures
and parallel methods are provided by the Portable Extensible Toolkit
for Scientific Computation \citep{petsc-web-page, petsc-user-ref}.

\section{Results of end-member models}

The theory described above is used to calculate grain size evolution
and two-phase flow in two end-member cases.  The first case is a
\textit{base model} with Newtonian rheology and spatially uniform
grain-size.  In the second, the \textit{full model}, we solve for a
fully coupled system with non-Newtonian rheology and dynamic
grain-size evolution.  Below we summarize the results of both sets of
simulations and their predictions for melt migration beneath the ridge
axis.

\subsection{Newtonian; constant grain size (base model)}

\begin{figure}[pt]
  \centering
  \includegraphics[width=0.8\textwidth]{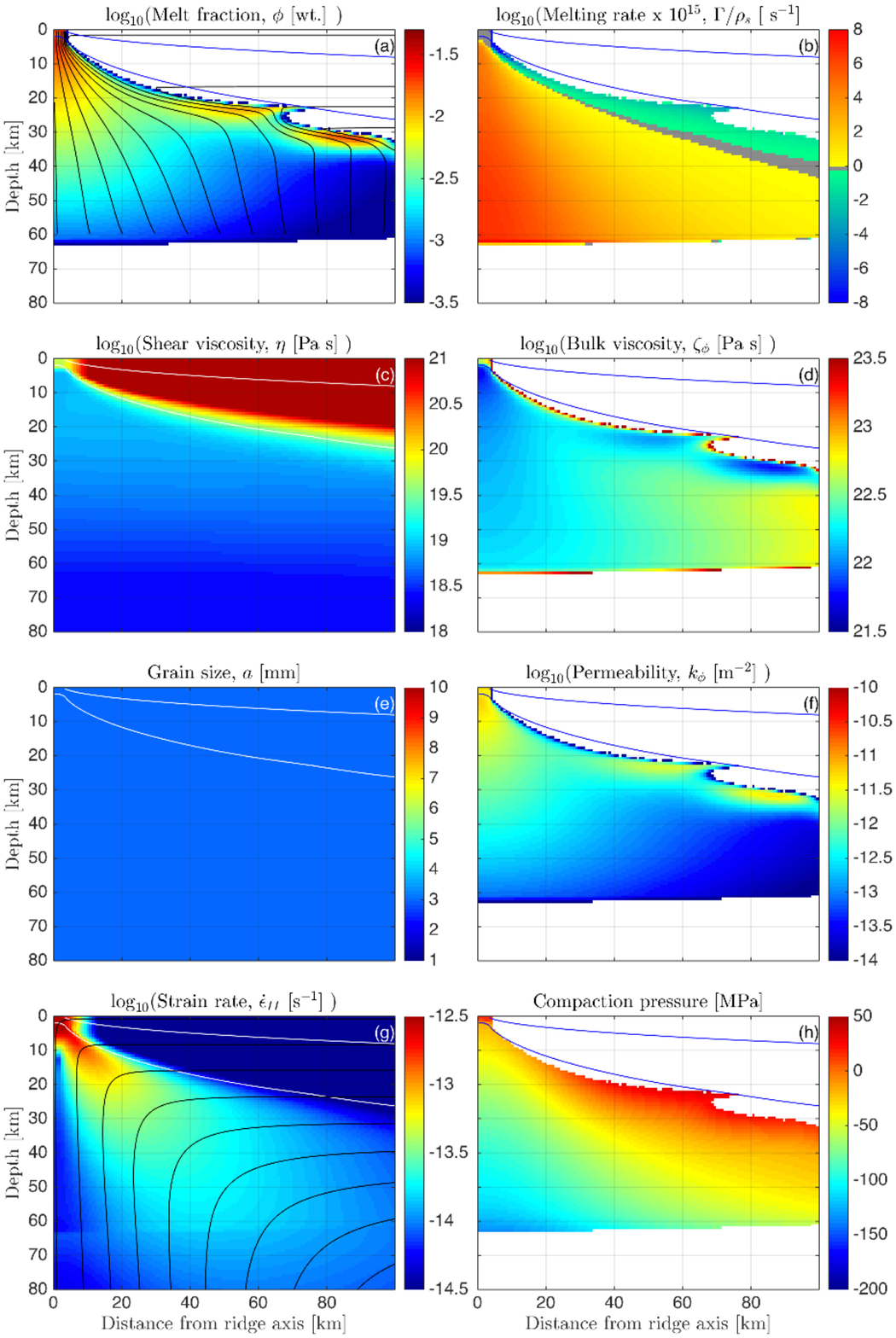}
  \caption{The base model run to a steady state (except for the
    compaction waves at the distal end of the melting region). The
    viscosity is Newtonian (eqn.~\eqref{eq:shear_viscosity}) with
    $n=1$, $m=0$. Each panel shows isotherms at 600 and 1250$^\circ$C
    as either white or blue lines. \textbf{(a)} Base-10 logarithm of
    melt fraction $\phi$ with streamlines of the magma flow-field
    $\vl$ (black) superimposed. \textbf{(b)} Volumetric
    volume-production rate of magma in inverse seconds, scaled by
    $10^{15}$. \textbf{(c)} Base-10 logarithm of shear viscosity
    $\eta$.  \textbf{(d)} Base-10 logarithm of compaction
    viscosity $\zeta_\phi$.  \textbf{(e)} Grain size in millimeters
    (constant and uniform, in this case). \textbf{(f)} Base-10
    logarithm of permeability in m$^2$.}
  \label{fig:base_model}
\end{figure}

The base model is constructed by imposing a static and spatially
uniform grain size of 3~mm.  Mantle rheology is constrained to be
Newtonian ($n = 1$) and grain-size insensitive ($m = 0$).  In this
case, the shear viscosity and permeability are solely functions of the 
temperature and melt fraction, respectively.  This model setup is
similar to that originally proposed by \cite{Sparks91} and calculated 
numerically by \cite{Katz10}.  

Melt migration in the base model is sub-vertical throughout much of
the melting region (Fig.~\ref{fig:base_model}a).  Immediately below
the lithospheric thermal boundary layer, melt is deflected toward the
ridge axis in a narrow zone, $\sim$5~km wide.  Melt fractions in the
mantle are generally low ($\le1$\%), increasing gradually upward and
toward the ridge axis. The location of the zone in which melt flow is
deflected horizontally is determine by the balance between the
compaction length, the rate of melt supply from below, and
crystallization from above \citep{Spiegelman93b}. The depth at which
the crystallization rate exceeds the melt supply closely follows the
1250$^\circ$C isotherm (Fig.~\ref{fig:base_model}b).  The deflection
of melt along this boundary results from the vertical gradient in
compaction pressure, which produces a net force that retards the
upward, buoyancy-driven transport of melt and deflects it laterally
toward the ridge axis.  This result is consistent with the
decompaction channel model originally proposed by \cite{Sparks91},
however, our calculations predict lower melt fractions in the channel.
Two possible reasons for this discrepancy are, first, a longer
compaction length in our models and, second, that our boundary
condition for melt extraction at the ridge axis adds an additional
component of suction toward the ridge.

Toward the margins of the melting region, solitary waves are observed
to arrive along the base of thermal boundary layer.  These features
are suppressed near the ridge axis due to the higher melting rates
(Fig.~\ref{fig:base_model}b) associated with enhanced mantle upwelling
directly beneath the axis \citep{Spiegelman93}.  On the limbs of the
partially molten region, however, the melting rate is insufficient to
suppress the formation of solitary waves, leading to a time-dependence
in the solutions, but only at distances $\gtrsim$50~km from the ridge
axis.

\subsection{Non-Newtonian; variable grain size (full model)}

\begin{figure}[pt]
  \centering
  \includegraphics[width=0.8\textwidth]{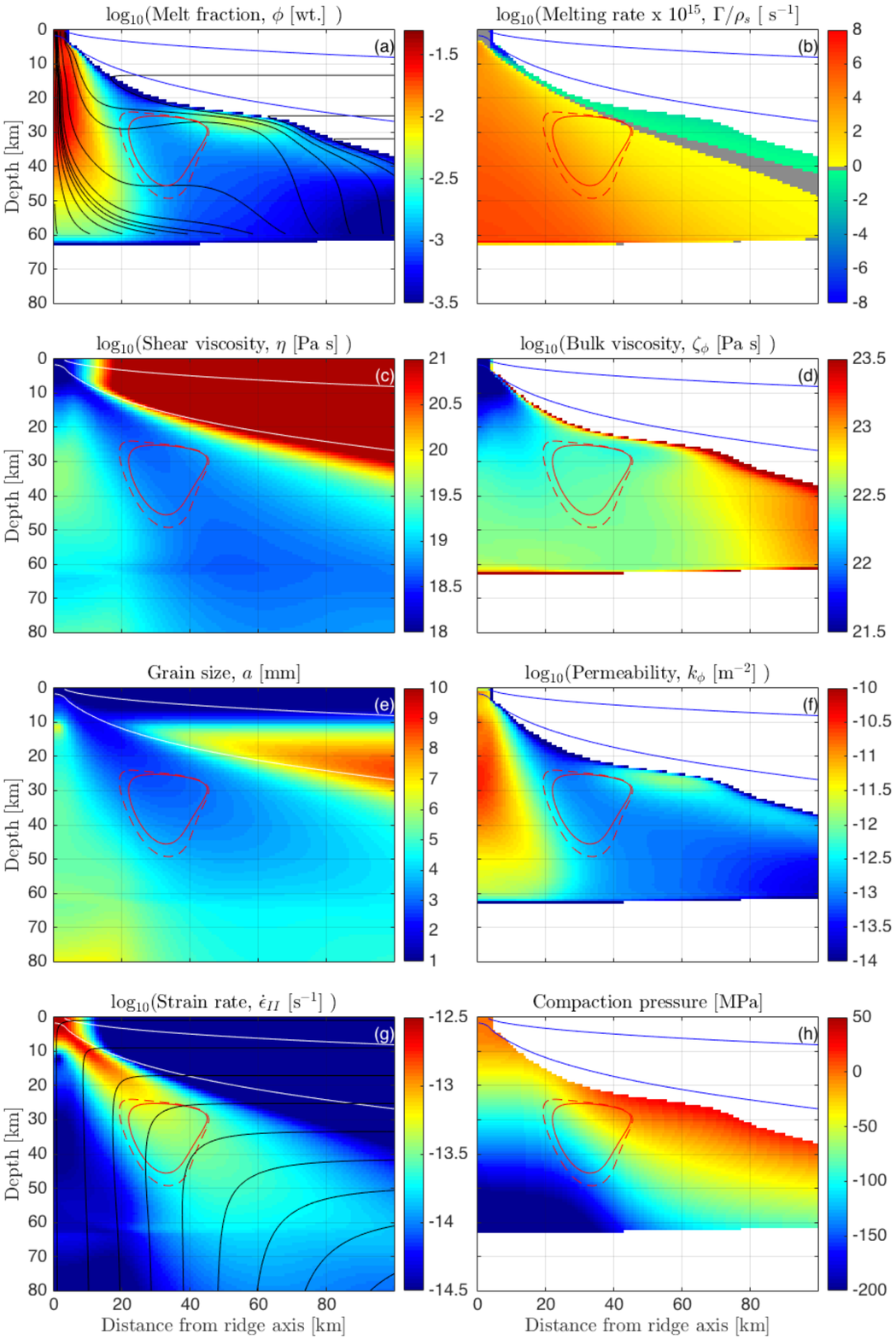}
  \caption{The full model run to a steady state. The viscosity is
    non-Newtonian (eqn.~\eqref{eq:shear_viscosity}) with $n=3.5$,
    $m=2$. Panels and isotherms as in Figure~\ref{fig:base_model}. The
    red lines are contours of the vertical component of the liquid
    velocity $\vl\cdot\zhat$. In particular, the dashed contour
    represents $\left(\vl-\vs\right)\cdot\zhat = 0$ and the solid
    contour represents $\left(\vl+\vs\right)\cdot\zhat = 0$. Within
    the region defined by the latter, melt flow is downward.}
  \label{fig:full_model}
\end{figure}

We next considered the full model in which grain size evolution is
coupled to the mantle deformation field (Fig.~\ref{fig:full_model}).
Mantle viscosity is non-Newtonian with a stress exponent ($n = 3.5$)
and grain size sensitivity ($m = 2$).  The resulting grain size field
is characterized by relatively large grain sizes directly beneath the
ridge axis and within the lower lithosphere
(Fig.~\ref{fig:full_model}e).  Grain sizes are smallest near the
surface and within a portion of the melting region that extends
downward from the ridge axis at an angle of $~45^\circ$.  This grain
size distribution is consistent with earlier single-phase simulations
that did not incorporate melt migration \citep{Turner15}. The dipping
layer of reduced grain size is the result of high strain-rates caused
by the corner flow; it corresponds to a region of low permeability
(Fig.~\ref{fig:full_model}e) that is not observed in the base model
(Fig.~\ref{fig:base_model}e). However, the variability in permeability
more closely mirrors the porosity
(Fig.~\ref{fig:full_model}a). Establishing causality, in this context,
appears to be a ``chicken/egg'' problem, but an analysis in the next
section clarifies the minimal importance of the permeability in the
overall dynamics.

The distribution of melt beneath the axis (Fig.~\ref{fig:full_model}a)
is dramatically different in the full model as compared to the
base model.  Specifically, melt is concentrated in a relatively narrow
region ($\sim$20~km wide) directly beneath the ridge axis, while the
melt fraction in the off-axis mantle decreases relative to the base
model.  These variations in melt fraction correspond to changes in the
trajectories of melt transport.  In the fully coupled model, melt flow
is directed vertically beneath the ridge axis, where melt fractions
are highest.  However, at distances $\gtrsim 10$~km from the ridge,
melt flow is sharply deflected towards the axis and follows horizontal
trajectories.  Indeed, at distances $\sim$20--40 km from the axis,
melts are predicted to travel slightly downward as they approach the
ridge axis (Fig.~\ref{fig:full_model}a).  A small zone of high melt
fraction is observed 40--70~km off-axis at the base of the thermal
boundary layer, but it is not a continuous feature that connects to the
ridge axis, as would be expected for a decompaction channel.

\section{Model sensitivity}
\label{sec:model-sensitivity}

The full model end-member shown in Figure~\ref{fig:full_model}
illustrates a striking difference in melt transport from the canonical
view associated with the base model, which has Newtonian viscosity and
constant grain size (Fig.~\ref{fig:base_model}).  These two
calculations differ in several ways: the permeability and the
viscosity structure associated with grain-size variation, as well as
the viscosity structure associated with the strain-rate
dependence. \cite{Turner15} considered that only the permeability
structure would modify melt segregation, but from these end-member
models, it is impossible to disentangle the contributions of these
different factors.

To analyse the role of each factor, we focus attention on the
sub-domain at lateral distances of 20 and 40~km from the ridge axis at
depths less than 60~km. There, the segregation of melt in the full
model is dominantly lateral, indicating that buoyancy is not the
primary driver of segregation. In fact there is a region, enclosed by
a dashed line, in which the segregation flux $\dflux\cdot\zhat \le 0$,
meaning that the vertical component of the liquid velocity is less
than that of the solid, though it may still be positive.  Within that
region, there is a smaller region in which $(\vl+\vs)\cdot\zhat \le
0$, which means that melt is driven downward, against
buoyancy. Downward melt flow was also obtained by \cite{Spiegelman87},
though in that case it was driven by the large dynamic pressure
gradients of unrealistically high-viscosity corner flow. Under these
conditions, where buoyancy is balanced by a vertical pressure
gradient, lateral melt transport can become dominant. Hence the
problem of understanding focusing can be approximately reduced to
understanding the vertical pressure gradient that balances buoyancy.

To this end it is helpful to consider a scaling analysis of governing
equations~\eqref{eq:governing_mechanics} in the $\zhat$
direction. Starting with the force balance of the aggregate,
eqn.~\eqref{eq:stokes}, we neglect body forces that are of order
$\phi$ and pressure gradients arising from corner flow, which are
small at these distances and at small shear viscosity
\citep{Spiegelman87}. This gives us $P \sim \zeta_\phi\Div\vs$, where
we have inferred that because the porosity is small,
$\zeta_\phi\gg\eta$. Then, taking $\omp\sim1$, expanding conservation
of mass eqn.~\eqref{eq:solid_mass} at steady state, and using it to
eliminate $\Div\vs$, we find that
\begin{equation}
  \label{eq:pressure_scaling}
  P \sim -\zeta_\phi \left(\Gamma/\rho - \vs\cdot\Grad\phi\right).
\end{equation}
This states that dynamic pressure is associated with the compaction
needed to balance melting and maintain constant porosity.

Substituting \eqref{eq:pressure_scaling} into eqn.~\eqref{eq:darcy} and
rearranging gives
$-\mobility\dflux \sim \Grad\left[\zeta_\phi\left(\vs\cdot\Grad\phi -
    \Gamma/\rho\right)\right] + \Delta\rho\gvec$. The case we are
interested in is $\dflux\cdot\zhat\le 0$ and hence we have
$\Delta\rho g \lesssim \diff{}{z}\left[\zeta_\phi\left(\vs\cdot\Grad\phi -
    \Gamma/\rho\right)\right]$.  The first term on the right-hand side
represents advection of porosity by the solid flow. This flow is
dominantly upward beneath the lithosphere in the region of
interest. Moreover, because melt extraction is efficient, the vertical
advection of porosity is much smaller than the production of melt
\citep[e.g.,][]{Ribe85}.  Hence, within the zone of negative melt
segregation ($\dflux\cdot\zhat\le 0$), it must be the case that
\begin{equation}
  \label{eq:scaling_balance}
  \Delta\rho g \lesssim - \frac{1}{\rho}\diff{}{z}\zeta_\phi\Gamma.
\end{equation}
Term-by-term evaluation of the numerical solution for the full model
confirms that this approximate balance holds.

It is evident from this analysis that changes in permeability cannot
balance the buoyancy of the magma.  The permeability (in terms of
mobility, $\mobility$) modulates the segregation flux but, because
permeability is scalar and strictly positive, it cannot explain the
downward segregation of magma. When the porosity is in steady state,
compaction must be in approximate balance with melting
($\Div\vs\sim-\Gamma/\rho$); this means that the dynamic pressure
scales as $P\sim-\zeta_\phi\Gamma/\rho$.  This pressure is typically
referred to as the \textit{compaction pressure} and its gradient can
drive magmatic segregation.  Equation~\eqref{eq:scaling_balance} shows
that compaction-pressure gradients, arising from melting-rate
variations, can drive segregation downward, against buoyancy. But the
melting-rate field, shown for each of the end-member models in
panel~(b) of Figures~\ref{fig:base_model} and \ref{fig:full_model}, is
nearly identical. Therefore the melting rate by itself cannot explain
the difference in segregation between these two models.

Covariation of the melting rate and the compaction viscosity
$\zeta_\phi$ can explain this difference. In particular, this
variation can explain the emergence of lateral melt transport in the
full model. $\zeta_\phi$ is the viscous resistance to compaction and
is shown in panel~(d) of Figures~\ref{fig:base_model} and
\ref{fig:full_model}.  In the base model, $\zeta_\phi$ increases
upward where $\Gamma$ decreases upward, leading to a product
$\zeta_\phi\Gamma$ that is roughly constant. In the full model, both
$\zeta_\phi$ and $\Gamma$ decrease upwards. The product
$\zeta_\phi\Gamma$ is thus not constant; instead it has a vertical
gradient that is sufficiently negative that the
inequality~\eqref{eq:scaling_balance} is satisfied.

Hence the crucial distinction between the base model and the full
model is not the grain-size control on melt mobility $\mobility$, as
was hypothesised by \cite{Turner15}. Rather it is the variation in
compaction viscosity $\zeta_\phi$ through its dependence on grain size
and strain rate, which it inherits from the shear viscosity
$\eta$. The grain size itself is sensitive to the strain rate; indeed
the rate of grain-size reduction scales with the square of the strain
rate. This raises the question of whether the strain-rate field,
through its influence on viscosity and grain size, can explain melt
focusing in the full model.  \cite{Rudge15} showed that if there is an
instantaneous balance of the rates of grain growth and reduction on
the right-hand side of eqn.~\eqref{eq:grain-size}, then the mean grain
size is given by a power law of the strain rate.  This can be
substituted into eqn.~\eqref{eq:shear_viscosity} for shear viscosity,
and hence the grain-size dependence can be combined with the
strain-rate dependence into a single power-law with an exponent
$n_e \equiv [n(p + 1) + m]/(p - m + 1)$. We will refer to $n_e$ as the
effective stress exponent.

Assuming reasonable values for the stress exponent $n$, grain rheology
exponent $m$, and grain growth exponent $p$ \citep{Karato89, Hirth03},
the effective stress exponents for dislocation, diffusion, and
grain-boundary sliding are calculated to be 3.5, 7, and 8,
respectively ($n_e=5.75$ in the Full model, above in
Fig.~\ref{fig:full_model}). A similar analysis was used by
\cite{Hansen12} to obtain an empirical flow law for deformation in the
grain-boundary sliding regime.  They obtained an effective stress
exponent of $n_e\sim$4.  These differences are likely related to
uncertainities in the grain growth exponent $p$ and grain rheology
exponent $m$.  Moreover, in natural systems where deformation is
accommodated by a composite of deformation mechanisms, $n_e$ will take
on an intermediate value.

\begin{figure}[pt]
  \centering
  \includegraphics[width=0.7\textwidth]{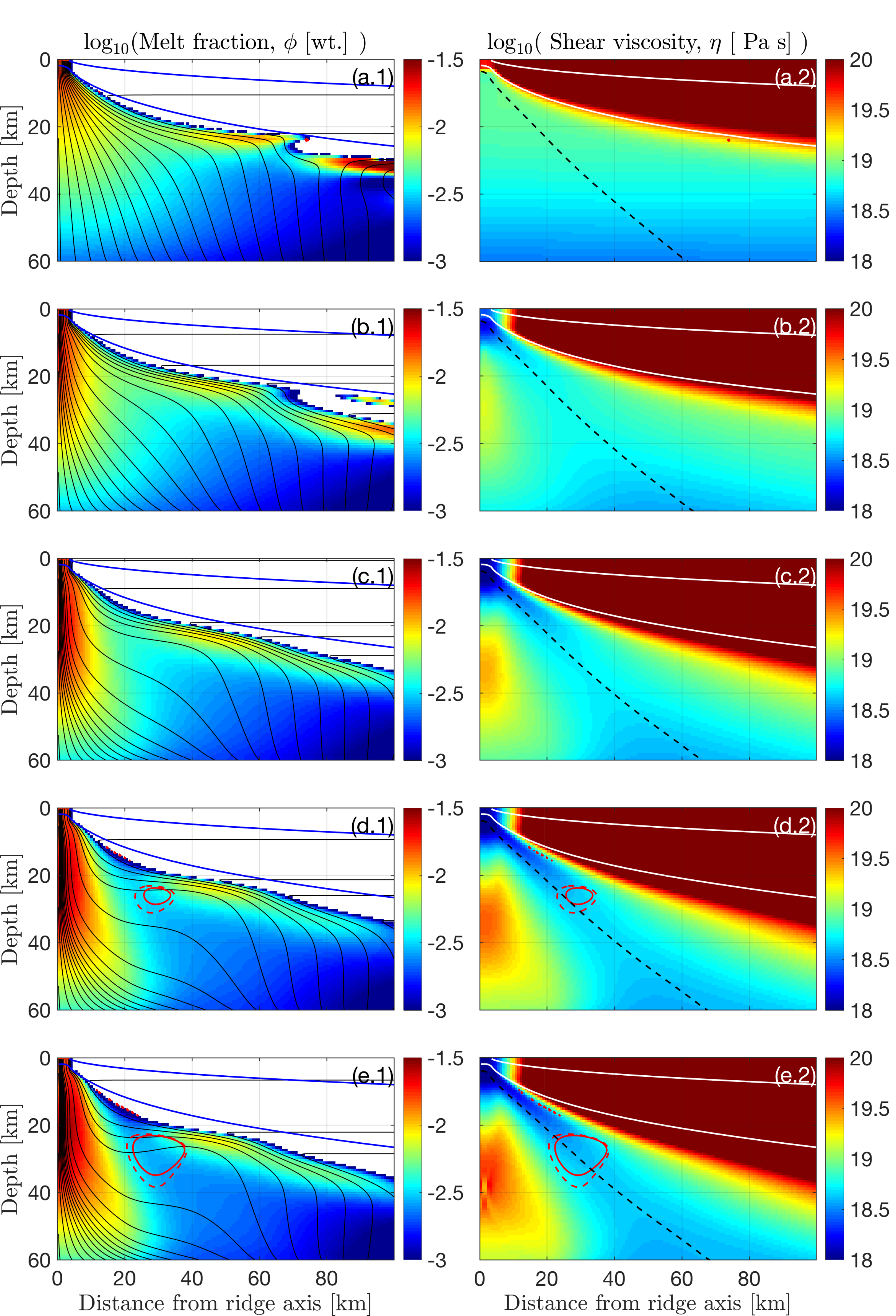}
  \caption{Model sensitivity to the effective stress exponent
    $n_e$. Rows show simulation output at steady state for
    $n_e=1,\,2,\,3.5,\,5.75,$ and $8$. Isotherms at 600 and
    1250$^\circ$C are shown in all panels with blue or white
    lines. Red lines are contours on which $\dflux\cdot\zhat = 0$
    (dashed) and $\vl\cdot\zhat = -\vs\cdot\zhat$ (solid). The first
    column shows the base-10 logarithm of melt fraction with
    streamlines of magma velocity $\vl$ superimposed. The second
    column shows the base-10 logarithm of shear viscosity
    $\eta$. The dashed line connects points where
    $\vs\cdot(\xhat-\zhat) = 0$; i.e., where the solid corner flow
    transitions from upwelling to lateral flow.}
  \label{fig:effective_exponent}
\end{figure}

Figure~\ref{fig:effective_exponent} shows output from simulations with
different values of the effective stress exponent, run to steady
state.  The grain size is constant and uniform in these models, and
hence does not affect the permeability or viscosity structure.  The
top row of panels has $n_e=1$, making it identical to the base model
shown in Fig.~\ref{fig:base_model}. Subsequent rows have
$n_e=2,\,3.5,\,5.75,$ and $8$. These values for $n_e$ are based on
experimental constraints on the viscosity parameters $n,\,m$ and the
grain-growth exponent $p$ for the dominant deformation mechanisms
in olivine, and thus represent the sensitivty of the shear viscosity
to stress or strain rate.

The shear viscosity field is shown in the second column of
Figure~\ref{fig:effective_exponent}. With increasing $n_e$, a
low-viscosity layer dipping at about $45^\circ$ emerges. The position
of this layer is set by the kinematic structure of corner flow;
namely, where the solid turns from upwelling to lateral motion there
is a larger strain rate that reduces the viscosity. This variation in
$\eta$ is mirrored in the compaction viscosity.

The first column in Fig.~\ref{fig:effective_exponent} shows porosity
and magma streamlines for increasing values of $n_e$. At larger
effective stress exponent, viscosity variations give rise to
compaction pressure gradients (eqn.~\eqref{eq:scaling_balance}) that
focus melt. The pattern of melt transport looks increasingly similar
to the pattern in the full model of Fig.~\ref{fig:full_model}, even
resulting in a region of downward melt segregation.  This demonstrates
that the difference between the full model and the base model is a
consequence of the strain-rate control (or, equivalently, the stress
control) on viscosity and not the influence of grain size on
permeability.

This finding is interesting, in part, because it is in stark contrast
to the invariance of the solid flow under changes to the effective
stress exponent.  This invariance is evident in the comparison of the
melting-rate field in panel~(b) of Figs.~\ref{fig:base_model} and
\ref{fig:full_model}, which reflects the solid upwelling rate.  The
solid flow is tightly constrained by the kinematic boundary conditions
and the thermally imposed structure of the lithosphere.  The liquid
flow, however, is free to respond to pressure gradients arising from
the dynamics and, in this case, from the compaction pressure
associated with melt extraction. The connection between melting,
compaction viscosity, and compaction pressure that was highlighted in
the scaling analysis, above, is the crucial link between rheology and
melt extraction. 

%Mantle viscosity is the result of a complicated interplay between different deformation mechanisms. 

% For the case of the shear viscosity $\eta$, these mechanisms are
% well studied by deformation experiments
% \citep{Kohlstedt15}. Compaction viscosity $\zeta_\phi$, by contrast,
% is poorly understood, though its consequences for melt transport are
% significant \citep{Alisic14, Alisic16}.

\section{Discussion and conclusion}

The results presented above represent a previously unexamined form of
magmatic focusing at mid-ocean ridges.  This mechanism does not rely
on the dynamic pressure gradient associated with corner-flow
\citep{Spiegelman87} or melt transport along a sublithospheric channel
\citep{Sparks91}, though both of these mechanisms can and do play a
minor role in the full model shown in
Figure~\ref{fig:full_model}. Instead, this focusing mechanism relies
on gradients in compaction pressure that arise from variations in
viscosity and melting rate.  These variations are, in turn, controlled
by the characteristic corner-flow structure of mantle deformation
beneath mid-ocean ridges.  To zeroth order, this flow structure is
controlled by the boundary conditions associated with plate separation
and is insensitive to viscosity variations.  Specifically, corner-flow
leads to a triangular region of upwelling with its apex at the
mid-ocean ridge axis.  This upwelling controls the melting rate field
$\Gamma$. The transition from upwelling to horizontal flow occurs
along the sides of the triangle, which slope downward away from the
ridge axis at an angle of $\sim$45$^\circ$. The enhanced strain rate
associated with this transition controls both the variation in
grain size $a$ and non-Newtonian viscosity $\eta$ (and
$\zeta_\phi$).

\begin{figure}[ht]
  \centering
  \includegraphics[width=0.9\textwidth]{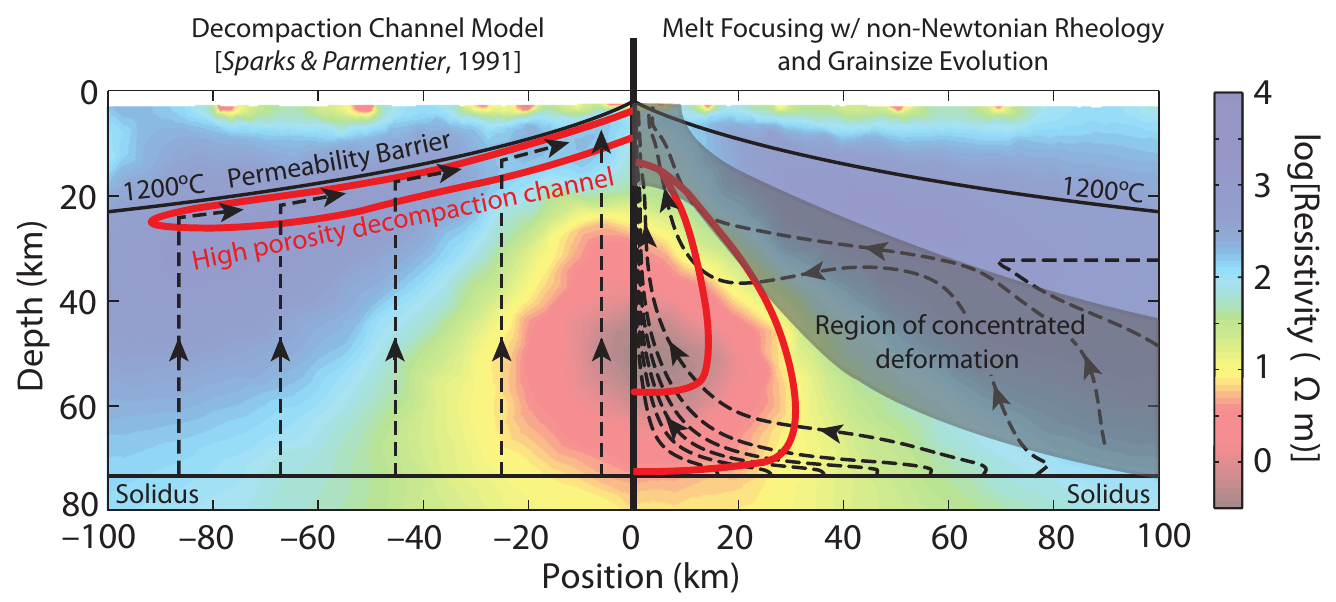}
  \caption{Schematic diagram of two hypotheses for melt focusing. The
    background image is the resistivity structure of the northern East
    Pacific Rise at $9^\circ30'$~N as obtained by \cite{Key13}. The
    magma streamlines on the left side of the diagram represent the
    focusing hypothesis of \cite{Sparks91}. Streamlines on the right
    side of the diagram are derived from the full model
    (Fig.~\ref{fig:full_model}) and depict the hypothesis proposed
    here. Red curves on both sides outline the region of increased
    melt fraction.}
  \label{fig:schematic}
\end{figure}

In the full model, the melting rate increases toward the ridge axis,
while the bulk viscosity remains roughly constant.  This results in a
spatial structure of $\zeta_\phi\Gamma$ that drives melt flow
laterally from the flanks of the melting region toward the ridge.
This can be understood by considering the balance of compaction and
melting when the melt fraction is constant.  At this steady state,
melt must segregate as fast as it is produced (neglecting the
advection of melt fraction, which is small). In other words, the solid
phase must converge at a rate that is given by the volume production
rate of magma, $-\Div\vs\sim\Gamma/\rho$. A convergent solid flow
means the magma is underpressured, i.e., that the compaction pressure
is negative.  This underpressure increases in magnitude with the
compaction viscosity.  Hence regions with slower melting rates and/or
lower viscosities have larger (less negative) compaction pressure. The
resulting compaction pressure gradient drives melt toward regions of
larger $\zeta_\phi\Gamma$.

The compaction pressure gradient associated with $\zeta_\phi\Gamma$ 
in the full model is enhanced by the focusing process itself. Focusing
concentrates melt into a column of mantle directly beneath the ridge
axis. The larger melt fractions in this column produce a
high-permeability connection between the surface and the melting
region below.  High permeability promotes a larger compaction length,
and hence the transmission of low compaction pressure from the ridge
axis to depth.

Figure~\ref{fig:schematic} is a schematic illustration of our
hypothesised focusing mechanism compared to the sublithospheric
decompaction-channel mechanism \citep{Sparks91}.  The resistivity
field obtained by \cite{Key13} is shown in the background.  Focusing
associated with a sublithospheric channel predicts a distribution of
melt fraction that is inconsistent with that implied by the
resistivity data.  In particular, the rapid, vertical extraction of
melt below the decompaction channel should result in higher and more
uniform resistivity throughout the melting region. Thus this model
cannot explain the steep-sided shape of the low resistivity region
beneath the axis.  By contrast, the full model predicts magma to move
laterally at depths above the onset of silicate melting and
concentrate beneath the ridge axis (Fig.~\ref{fig:schematic}).  This
melt distribution is more consistent, qualitatively, with the MT
inversion, which shows a minimum in resisitivity directly beneath the
ridge axis.

The focusing mechanism we propose is different in many of its details
from that suggested by \cite{Turner15}. Their hypothesis was based on
the permeability structure arising from variations in mean grain
size. The grain size variations obtained in the full model above
(Fig.~\ref{fig:full_model}) are largely consistent with
\cite{Turner15}. However, our numerical experiments using constant
grain size and large $n_e$ in section~\ref{sec:model-sensitivity} show
that focusing is not a direct consequence of the permeability
structure. Furthermore, the scaling analysis presented above supports
the idea that viscosity and melting-rate variations are most important
in promoting melt focusing.  Thus, while grain size can be important
in influencing mantle viscosity, its effect on mantle permeability
does not exert a significant control on melt focusing.

This conclusion raises a more general point regarding two-phase models
of coupled magma/mantle flow. For convenience and simplicity, those
models have generally neglected the non-Newtonian stress dependence
that arises from dislocation-based creep mechanisms when computing
two-phase flow \citep[e.g.,][]{Katz10}. This has been justified on the
grounds that the kinematically-driven mantle flow field is little
affected by this exclusion. However, the results presented here
suggest that despite the invariance of the solid flow, the pattern of
magmatic segregation is sensitive to the stress dependence of mantle
viscosity. Although this has been noted in models of laboratory-scale
two-phase flow \citep{Katz06}, it was unrecognised in the context of
larger-scale tectonic models.  Future work should more thoroughly
explore the consequences of non-Newtonian viscosity for melt
segregation in different geologic environments.

\paragraph{Acknowledgements} The research leading to these results has
received funding from the European Research Council under the European
Union’s Seventh Framework Programme (FP7/2007–2013)/ERC grant
agreement 279925. The authors acknowledge the use of the University of
Oxford Advanced Research Computing (ARC) facility in carrying out this
work. Support for Behn was provided by grant OCE-14-58201 from the US
National Science Foundation. The authors thank the Isaac Newton
Institute for Mathematical Sciences for its hospitality during the
programme Melt in the Mantle, which was supported by EPSRC Grant
Number EP/K032208/1.

\bibliographystyle{abbrvnat}
\bibliography{manuscript}
\end{document}